# Network Analysis in the Legal Domain: A complex model for European Union legal sources

Marios Koniaris[1], Ioannis Anagnostopoulos[2], and Yannis Vassiliou[1]

[1] KDBS Lab, School of ECE, National Technical University of Athens, Greece
[2] Department of Computer Science & Biomedical Inf., University of Thessaly, Greece

**Abstract.** Legislators, designers of legal information systems, as well as citizens face often problems due to the interdependence of the laws and the growing number of references needed to interpret them. In this paper, we introduce the "Legislation Network" as a novel approach to address several quite challenging issues for identifying and quantifying the complexity inside the Legal Domain. We have collected an extensive data set of a more than 60-year old legislation corpus, as published in the Official Journal of the European Union, and we further analysed it as a complex network, thus gaining insight into its topological structure. Among other issues, we have performed a temporal analysis of the evolution of the Legislation Network, as well as a robust resilience test to assess its vulnerability under specific cases that may lead to possible breakdowns. Results are quite promising, showing that our approach can lead towards an enhanced explanation in respect to the structure and evolution of legislation properties.

## 1 Introduction

Legislation is a large collection of different normative documents, which keeps growing and changing with time. As legislation increases in size and complexity, finding a relevant norm may be a challenging task even for experts.

Furthermore, the process of drawing up a consistent and coherent legislation framework becomes a more and more challenging task. Drafting of new or amending existing legislation are very complicated processes. As a result, authorities at European, national and local level, often consider proposed regulations for months or years before they finally become effective. Thus, it is critical to firstly quantify the legal complexity and then work towards the provision of a model that will assist us to reveal the emergent dependencies among the legislation corpus.

Typically, legal documents refer to authoritative documents and sources e.g., most commonly regulations, treaties, court decisions, and statutes. Computer scientists and legal experts have used citation analysis methods, in order to construct case law citation networks, as well as to further model and quantify the complexity of the legislation corpus [1,2,3,4].

However, studied networks contain only court decisions, making them less suitable for other legal systems than Common law e.g., Civil law. Also, relations between case documents on the studied networks are only references. Thus, the hierarchical structure of the normative system is absent from the adopted model. Our approach differs from previous works that deal with the specific problem, as we do not utilize the legislation corpus in terms of a citation network. Instead, we employ a multi relationship model in which two or more legal documents, belonging to the same or different types, may be linked to others by more than one relationships. Unlike previous studies of legal citation networks, our model encompasses many aspects such as hierarchy between the sources of law and the different types of relations between legal documents. This modeling approach transforms legislation corpus into a multi-relational network: a network with a heterogeneous set of edge labels that can represent relationships of various categories in a single data structure.

We investigate the topological structure of the Legislation Network to discover properties and behaviours that transcend the modeling abstraction. Results are quite promising, showing that the Legislation Network is a scale free, small-world network. This can be reflected as an evolutionary advantage since these kind of networks are more robust to disturbance than other network architectures [5].

Since the legislation corpus evolves over time, a temporal analysis of the evolution of the Legislation Network reveals otherwise hidden aspects of the legislation process. The Legislation Network is obeying densification power law [6], with the number of active edges, connections between legal documents, growing faster than the number of active node, legal documents.

We also performed a resilience test on the Legislation Network in order to understand and predict the behaviour of the network under malfunctions. We analysed its behaviour when its nodes (legal documents) or edges (connections) between them are removed. This may be the result of a temporal process since legislation evolves over time e.g,. law that is amended, invalidated or cease to exist.

In this paper, we propose a novel approach to model the legislation corpus. A model that can be applied to civil law collections, such as the laws of the European Union. To the best of our knowledge, our work is the first work that deals with the specific problem in a sense that it: i) models civil law as a network with various types of relations, ii) identifies several topological characteristics of it, iii) performs a temporal analysis over the evolution of the legislation corpus and iv) performs a resilience test on the legislation corpus to assess its vulnerability.

The rest of this paper is organized as follows. Section 2 briefly reviews related work and approaches. In Section 3 we provide a short overview of network analysis principles, introduce the examined datasets and our network construction method. In Section 4 we analyse the structure and the temporal evolution of the legislation corpus and perform a resilience test on it. Finally, Section 5 concludes and discusses further work.

## 2   Related work

Citation analysis has been used in the field of law to construct case law citation networks [7]. The American legal system has been the one that has undergone the widest series of studies in this direction. [8] examined the network structure of precedent-based judicial decision making, using data from United States Supreme Court. Fowler et al. [9] experimented with methods to identify the most central decisions of the US Supreme Court, while afterwards [1] they studied how the norm of stare decisis [3] [10] had changed over time in the jurisprudence of the US Supreme Court, as to identify the doctrine's most important related precedents [4].

In [11] the network of Canadian case law is examined with network analysis algorithms, concluding that indegree centrality and PageRank scores of case law network are effective predictors of the frequency with which those cases will be viewed on the Canadian Legal Information Institute website. In contrast, van Opijnen [12] concluded that network algorithms, which have been used in previous research, especially in-degree, HITS and PageRank [13], might not be the most appropriate to measure legal authority. The same researcher proposed a model for automated rating of Case law which incorporates data from the publication and the citation of legal cases to estimate the legal importance of judgments [14].

Smith observed that the network of US Supreme Court decisions followed a power-law distribution [3]. The authors of [15] described a visualization-based interactive legal research tool that allows users to easily navigate in the legal semantic citation networks and study how citations are interrelated. In [4] a framework for measuring relative legal complexity is proposed, taking into account the structure, language, and interdependence of legal sources.

However, these studies focus on Common law: a law developed by judges through decisions of courts, which is fundamentally different with the Civil law that is used across the European Union. For quantifying the complexity of the judicial corpus through network analysis in the Civil law domain, Winkels et al have used a sample of 15,053 cases from the Dutch Supreme Court [16]. The authors verified that Fowler results also apply for the citation network of the sampled Dutch legal system. Similarly, the complexity of the French legal code was analysed in the work described in [2], where the authors identified structural properties of the French legal code network by sampling 52 legal codes.

Precedent in international courts is studied in [17]. Authors applied network analysis techniques to case citations by the European court of human rights as to conclude that international and domestic review courts develop their authority in similar ways. In a analogous manner, the International Criminal Court [18], the Italian Constitutional Court [19] were examined from a network-science perspective. Finally, a toolkit allowing legal scholars to apply network and visual analytics techniques to E.U. case law is presented in [20].

In all of the above studies, the case law corpus is treated as a citation network, thus showing the effectiveness of network analysis in the legal domain. In one hand, it was supported that case law citation networks contain valuable information, capable of measuring legal authority, identifying authoritative precedent, evaluating the relevance of court decisions, or even predicting the cases that will receive more citations in the future. Yet, on the other hand, citation network analysis over the legislation corpus, provides us information over a single dimension view. Edges on the graph are of the same type and just simple references between judicial documents.

---

[3] A legal norm inherited from English common law that encourages judges to follow precedent by letting the past decision stand.

[4] A judicial decision in a court case that may serve as an authoritative example in future similar cases.

However, in the real-life paradigm of legal domain, there are multiple and heterogeneous networks, each representing a particular kind of relationship, and each kind of relationship plays a distinct role in a particular legal norm. Thus, in order to construct a network model that simulates legislation in a quite robust way, we have to take into account the multi-scale structure of law. Distinct features of the law as the hierarchy between the sources of law, or different types of relations between legal documents should be properly carved and incorporated into a model, as we further analyse in the following sections.

## 3   Legislation Modelling

In this work, we model the way laws are correlated through their graph properties. However, in order to fully model the legislation corpus we have to properly analyse it and carefully identify its unique features. In the following sections we discusses some background information on network analysis concepts and techniques (Section 3.1), describe the dataset used for the legislation analysis (Section 3.2) and the way the respective network is constructed (Section 3.3).

### 3.1   Background

In this section, we give a short introduction of relevant concepts and techniques proposed in the literature.

A network, also called graph, is a set of items, called vertices or edges, with connections between them, called arcs or edges. More formally, a graph is an ordered pair $G = (V, A)$ consisting of a set $V$ of vertices (nodes) with a set $A$ of arcs (edges), which are 2-element subsets of $V$. This rather simple modeling framework can be made more powerful if one extends it to include additional levels of detail. For example there may be more than one different types of nodes/edges, nodes and edges could appear and disappear within time periods, edges could be directed, pointing in only one direction, or not, multiple edges may exist between the same pair of nodes etc. Thus, graphs whose structure is irregular, complex and dynamically evolving in time can be formed and used to describe a wide variety of underlying systems.

Usually, networks are the infrastructure of some system, e.g., disease spreading on social networks, and this system is what we are really interested in. The advantage of modeling a system as a network is that we can say much about the behavior of the system without studying the actual dynamics at all. Analyzing the structure of a network, one can reveal important clues about its behavior, e.g., predict how fast a virus will spread [21], assess which are the most important nodes [22,13] and predict how robust to damage an area of the network is [23].

Various network models have been proposed in the literature to help us understand or even predict the behavior of natural or man-made systems e.g., transportation networks [24], the internet [25], food webs [26], the network of metabolic pathways [27] , networks of brain neurons [28], social networks [29] and many others.

Research has shown that real networks are far from being random, but display generic organizing principles [30]. Among the mathematical properties that characterize these principles are:

- **Degree distribution**. The degree of a node in a network is the number of edges the node has to other nodes. Since not all nodes in a network have the same number of edges, the degree distribution $P(k)$ of a network is defined to be the fraction of nodes in the network with degree $k$. In other words, the distribution function P(k), gives the probability that a randomly selected node has exactly $k$ edges.

  In random graphs [31], where edges are placed randomly, the majority of nodes have approximately the same degree, close to the average degree $k$ of the network and the degree distribution is a Poisson distribution.

  Unlike random graphs, for a large number of real networks, there are quite a few very highly connected nodes and the vast majority of nodes has only few connections to other nodes [5]. Since the range of degree values varies very greatly, such a network is called scale-free network. In many cases the degree distribution $P(k)$ follows a power law; that means that the way the probability decreases with $k$ seems to be a reasonably close fit to $K^{-\gamma}$ for some $\gamma$, i.e., $P(k) \infty K^{-\gamma}$. The parameter $\gamma$, called the power-law exponent for the degree distribution, varies between 2 and 3 in real-world networks, although it may lie outside these bounds.

  Such deviations from the normal distribution usually signify some important correlations within the system. Fat tailed probability distributions have been detected in many complex systems, spanning different branches of natural sciences, as well as social phenomena.

- **Clustering coefficient**. The clustering coefficient quantifies the tendency of nodes to cluster together. The clustering coefficient of a node is defined by the proportion of links (edges) between the nodes within its neighborhood, immediately connected nodes, divided by the number of links that could possibly exist between them.

  In real networks the clustering coefficient is typically much larger than it is in a random network of equal number of nodes and edges [5]. In practice, a high clustering coefficient, compared to random graph of same size, indicates that there are groups (clusters) of nodes that are highly interconnected among themselves, but have few connections to other clusters.

- **Small world (six degrees of separation)**. A small-world network [29,32] is a network where the number of steps required to travel between two randomly chosen nodes grows sufficiently slowly as a function of the number of nodes in the network. In other words, despite their often large size, there is a relatively short path between any two nodes. Thus, in small world networks, while most nodes are not neighbors of one another, most nodes can be reached from every other by a small number of steps. Small-world properties are found in many real-world networks, e.g., railway networks [24], metabolite processing networks [33], networks of brain neurons [28], food webs [26] and the World Wide Web [25].

- **Weakness in spite of overall strength**.
  An interesting phenomenon of complex networks is their "Achilles' heel": robustness versus fragility [34]. In a power law distributed small world network, deletion of a random node rarely causes a dramatic increase in shortest path length. By contrast, in a random network, in which all nodes have roughly the same number of connections, deleting a random node is likely to increase the mean-shortest path length slightly, but significantly, for almost any node deleted. In this sense, random networks are vulnerable to random perturbations, whereas small-world networks are robust. However, small-world networks are vulnerable to targeted attack of hubs, whereas random networks cannot be targeted for catastrophic failure [23,35].

### 3.2   Dataset used for the legislation analysis

European Union law consists of founding treaties and legislation, such as Regulations and Directives, which have direct or indirect effect on the laws of European Union member states. There are three sources of European Union (EU) law:

a **primary**, the Treaties establishing the EU,
b **secondary**, regulations and directives which are based on the Treaties,
c **supplementary law**, the case law of the Court of Justice, international law and the general principles of law.

The official legal portal of the European Communities is offered by the EUR-Lex[5], a free public service for the dissemination of EU law. EUR-Lex contains all documents printed in the Official Journal of the EU dating back to 1951. For the purposes of our work, we have downloaded all documents since then and we have extracted unnecessary html formatting option, in order to obtain a text copy of the European Communities legal database.

Within this database, documents are organized into sectors. Table 1 summarizes the sectors of the EUR-Lex database, along with their corresponding number of documents, as of July 2013. We have extracted all legislation concerning Sectors 1 to 6 from the database, in accordance with the three sources of EU law, accounting for a total number of 249,690 documents.

Table 1: Explanation of the EUR-Lex sector classification mechanism (#docs corresponds to the number of documents within each sector, as of July 2013)

| | EUR-Lex Classification | | |
|---|---|---|---|
| Sec. | Title | Explanation | # docs |
| 1 | Treaties | Treaties establishing the EU / supplementing Treaties | 8652 |
| 2 | International agreements | Agreements between the EU and other sovereign countries | 8564 |
| 3 | Legislation | Secondary legislation to implement EU policy | 120,550 |
| 4 | Complementary legislation | Agreements between Member States | 1231 |
| 5 | Preparatory acts | Proposals for future legislation/ opinions | 73,123 |
| 6 | Jurisprudence | Case law (judgments, orders, interpretations and other acts) | 37,570 |
| | | TOTAL | 249,690 |

EUR-Lex database offers analytical metadata for each document. The bibliographic notes of the documents contain information such as dates of effect and validity, the legal form of the document, authors, the subject matter, the legal document from which the document draws its authority, as well as various relationships to other documents and classifications.

We considered that fields, which provide links to other documents in the database, are of particular significance and importance for our study. In Figure 1, we provide a visual representation example of a sequence of modifications imposed to a legal document in the form of amendments. The council directive *370L0220*, dated 20 March 1970, was amended by directive *383L0351* in 16 June 1983 and then further amended by directive *389L0491* in 17 July 1989. Note that this is a bidirectional relationship, since directive *383L0351* modifies/amends directive *370L0220* then directive *370L0220* is amended by *383L0351*.

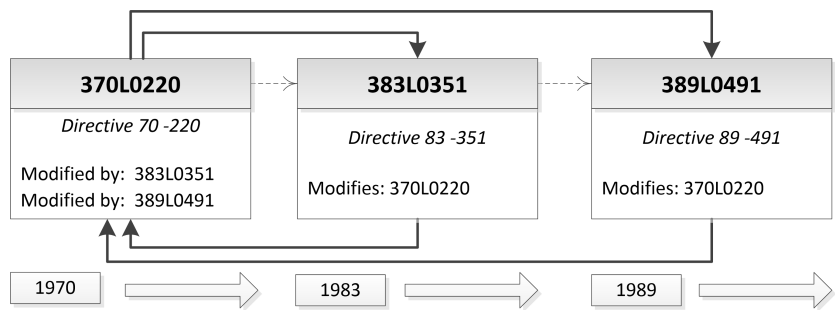

Fig. 1: Cross-reference links between legal documents in the EUR-lex. Amended by and Amendment to are bidirectional relationships

References in the legislation can be divided into two different categories[6] : (a) read only references that do not modify the target document and (b) edit references that modify either the text or the lifecycle of the target document. Instruments cited is an example of the former, while amended by is an example of the latter.

Table 2 provides an overview of the major category types for the references found in the EU law database. It also identifies that the "Instruments cited" reference type consists of more than the half of the Legislation Network (close to 55%). If we consider the respective corpus as simple instances of citation networks, like previous studies, then we would have nearly neglected 45% of the total relations. This also indicates that previous studies, that focus solely on citation analysis over legal corpora, ignore a significant amount of the networks properties.

---

[6] Internal reference is a reference that points to an article in the same regulation and is excluded from the scope of our study.

Table 2: Type of references found in the EUR-Lex

| EUR-Lex cross-references | | |
|---|---|---|
| Type | Explanation | % of Ref |
| Amended by | The document is amended by another | 9,50% |
| Amendment to | The document amends another document | 9,50% |
| Legal basis | The document is authorized by the mentioning document | 23,50% |
| Instruments cited | The document cites other docs | 54,93% |
| Affected by case | The document was altered as of a case result | 2,00% |
| Other | Various types of references. | 0,57% |

### 3.3  Legislation Network Construction

Generally, legislation consists of a number of normative documents that are cross-referred to each other. Thus, a directed network can be formed if a legal document refers to another (outgoing link), or is refereed by another document (incoming link). Furthermore, since legal documents can only refer to existing ones, our modeling graph is in fact a directed acyclic graph (DAG).

Figure 2 displays the formation of EU law network from the legal document database. Nodes of the network represent the legal documents. Every document in the legal collection is analysed for cross references. If a cross reference is found between two documents, then a suitable edge connects those two nodes.

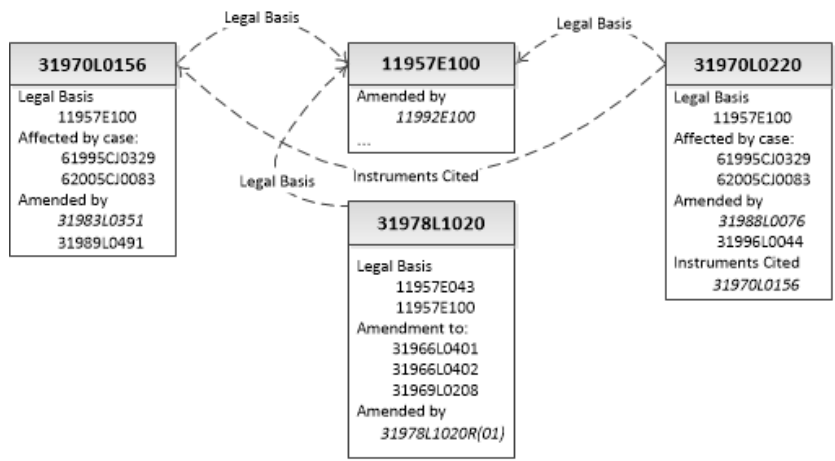

Fig. 2: A fraction of the EU Law Network

Node types vary according to the corresponding sector of the legal document, as already explained in Table 1. Edges of the graph have many types according to the types of references found in the EUR-Lex database, as depicted in Table 2. In total the graph consists of 234,287 nodes and 998,595 edges connecting the nodes.

Nodes and edges on the legislation network have temporal attributes also. Each node is marked with a *date of effect*, the date that the legislation became effective and a *date of expiry*, the date that the legislation will cease to effect. Quite often legislation is adopted without an explicitly stated expiry date, also called as *sunset close*. For those nodes, without a sunset close, we have set an expiration date for the year 9999. Edges follow the temporal distribution of the corresponding nodes. That is, an edge is considered valid only for the time periods between the effective dates and sunset close of the nodes they connect. This characteristic attribute of the Legislation Network is of special importance; it allows to reproduce active legislation in any given point in time.

The EU Legislation Network, as many real-world networks, exhibits both temporal evolution and multi-scale structure. It is a multilayer network [36], as it is a network with a heterogeneous set of edge labels, which represent references of various types (Legal basis, Instruments cited, etc.).

Figure 3 provides a visual representation of the Legislation Network. Various legal documents such as Treaties (red nodes/ layer), International agreements (blue nodes/ layer) and Legislation (green nodes/ layer) are connected through edges of type Legal Basis (dotted line) and Instruments cited (continuous line).

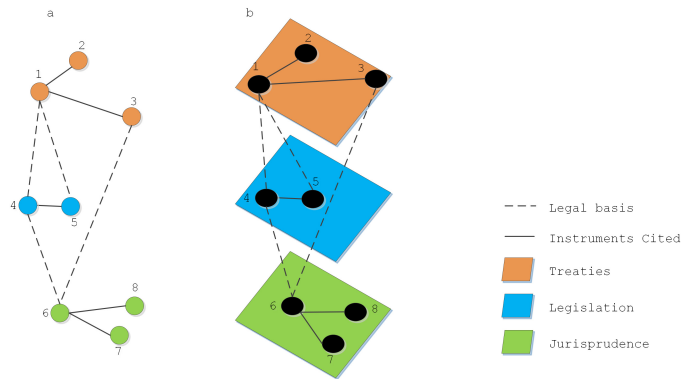

Fig. 3: (a) An example of the multi layer structure of the Legislation Network. Legal documents belonging to different sectors, represented with different colors, are interconnected with different types of relationships i.e., Legal Basis, Instruments cited. (b) Representation of the same network using layers. (Best viewed in color.)

Alongside with the whole network, **Legislation Network (LN)**, we identified the following sub networks, which we additionally examine in detail through the rest of the paper :

– The (sub) network of **Regulations (RN)**. In this network, we keep track only of legal documents that belong to the sector Legislation of EUR-Lex. We identified this network as this network accounts for the corpus of secondary legislation to implement EU policy, as it contains EU regulations[7], directives [8] and decisions[9] with a direct or indirect effect on EU member states.

- The (sub) network of **Instruments cited (ICN)**. The network of Instruments cited contains all documents of the Legislation Network and only those edges connecting nodes of type Instruments cited. We identified this network as it resembles a citation network as it is studied in previous works
- The (sub) network of **Legal basis (LBN)**. Within this network, we keep only the edges of type Legal basis. An edge is added the network from node legal document A to node legal document B if A is authorized by B. This network is of great importance for everyone trying to identify the internal hierarchy of the legislation corpus.

Nevertheless, our approach is of general usage and any particular combination of node and edge filtering technique can be applied within the proposed modeling approach; a researcher interested only in Case law e.g., judgments from the court of Justice, may study the corresponding Jurisprudence Network, a researcher studying the evolution of E.U. treaties may confide his/her analysis on the Network of Treaties, while one interested in identifying the effects of court decisions over the legislation may partition the Legislation Network based on the "Affected by case" type of legal reference.

In order to construct the sub-networks we divide the Legislation Network into sub-graphs based on the following criteria: sector type, reference type, time period or even a combination of them. Algorithm 1, described below, divides the legislation graph in a sub-graph of specific sector of legislation. Corresponding E.U. legislation Sectors are presented in Table 1.

---

**Algorithm 1** Produce legislation graph of specific sector

---

**Input:** legislation graph $G$, legislation sector $s$
**Output:** legislation graph $G$ of specific sector
    $Sectors \leftarrow$ list of legislation sectors
    **for all** $sector \in Sectors$ **do**
        **if** $s \neq sector$ **then**
            $n \leftarrow$ nodes in $G$ of sector type $s$
            $e \leftarrow$ edges($n$)
            $G \leftarrow G \subset (n, e)$
        **end if**
    **end for**
    **return** $G$

---

Similarly, Algorithm 2 separates the legislation graph in a sub-graph of specific relations. Applicable types of legislation references are presented in Table 2.

---

[7] Regulations are of general application, binding in their entirety and directly applicable.

[8] Directives are binding, as to the result to be achieved, upon any or all of the Member States to whom they are addressed.

[9] Decisions are binding in their entirety.

---

**Algorithm 2** Produce legislation graph with specific relations

---

**Input:** legislation graph $G$, relation type $r$
**Output:** legislation graph $G$ of specific relation
 1: *Relations* ← list of legislation relations
 2: **for all** *relation* ∈ *Relations* **do**
 3:     **if** $r \neq relation$ **then**
 4:         $e$ ← edges in $G$ of relation type $r$
 5:         $G \leftarrow G \subset (e)$
 6:     **end if**
 7: **end for**
 8: **return** $G$

---

While access to legislation generally retrieves the current legislation on a topic, point-in-time legislation systems address a different problem, namely that lawyers, judges and anyone else considering the legal implications of past events need to know what the legislation stated at some point in the past when a transaction or events occurred which have led to a dispute and perhaps to litigation [37].

The following Algorithm 3 can be applied to create a sub graph that represent legislation in effect for a given time step frame.

---

**Algorithm 3** Produce legislation in effect for time period t

---

**Input:** Complete legislation graph $G$, time step period $t$
**Output:** legislation (in effect) graph $G$ for time period $t$
 1: $n1$ ← expired nodes in $G$                                    ▷ date of *expiration* $< t$
 2: $e1$ ← edges($n1$)
 3: $n2$ ← (future) nodes in $G$,                                  ▷ date of affect $> t$
 4: $e2$ ← edges($n2$)
 5: $G \leftarrow G \subset (n1, n2, e1, e2)$
 6: **return** $G$

---

Table 3 summarizes various properties of the Legislation Network and sub networks that we further analyze. For each network we indicate the number of nodes, the number of edges, the average degree, the diameter, the average path length, the size of the giant component (g.c.) and the number of isolated nodes. Additionally, using parentheses, we display the metrics for the current version of each sub network. That is, using Algorithm 3 we form the current (active) version of each network and measure the aforementioned properties.

The diameter and average path length of the Instruments cited network appear analogously higher compared with the other networks. In contrast, the average degree and average path length metrics of the Legal basis network are quite smaller than in the other networks, as this network contains only edges of type Legal Basis. Furthermore, we note that the current (active) version of each sub-network is less connected, a smaller percent of nodes belong to the g.c. and consequently there are more isolated nodes.

Table 3: Legislation Network and sub Networks basic properties. Numbers in parentheses refer to the current (active) version of each network.

| Metrics/ Network | Legislation (LN) | Regulations (RN) | Inst. cited (ICN) | Legal basis (LBN) |
|---|---|---|---|---|
| # of nodes | 234,287 (122,091) | 115,105 (36,330) | 140,208 (85,417) | 163,095 (51,898) |
| # of edges | 998,595 (524,503) | 338,134 (72,605) | 554,917 (387,803) | 237,531 (74,467) |
| Average degree | 8.52 (8.59) | 5.88 (4.00) | 7.92 (9.08) | 2.91 (2.87) |
| Network diameter | 39 (33) | 41 (30) | 79 (60) | 6 (6) |
| Average path length | 7.22 (7.58) | 7.00 (7.20) | 7.54 (6.90) | 1.66 (1:48) |
| Size of g.c. | 233,337 (116,790) | 112,532 (29,583) | 133,211 (78,140) | 161,081 (49,038) |
| % of g.c | 99.6% (95.7%) | 97.8% (81.4%) | 95% (91.5%) | 98.8% (94.5%) |
| Isolated nodes | 950 (5,301) | 2,573 (6,747) | 6,997 (7,277) | 2,014 (2,860) |

## 4    Network Analysis

In this section, we apply our modeling approach as to identify/characterize various properties of the Legislation network. We examine the Legislation Network structure and try to identify, otherwise, hidden organizing principles of the legislation corpus (Section 4.1). We proceed with studying how the legislation corpus evolves over time (Section 4.2), as new laws get introduced and others are amended or invalidated. Finally, we evaluate the tolerance of the Legislation Network to errors/ breakdowns, by performing a resilience test (Section 4.3). The Legislation Network structure was presented in a preliminary work of ours [38], while in this work, we enrich our model and extend the study by adding its Temporal Evolution and Resilience test.

### 4.1    Network Structure

The characterization of the structural properties of the underlying network is a very crucial issue to understand the function of a complex system [39]. An important realization of network analysis is that networks in natural, technological and social systems are not random, but follow a series of basic organizing principles in their structure and evolution, thus distinguishing them from randomly linked networks [5]. Network structure analysis inspects both macro and micro measures of the network topology. Macro measures describe the network structure in a global view and help us interpret the influence of network structure to individual nodes in the network. On the contrary, micro measures quantify the relative importance of a node within the network and assist us perceive the influence of individual nodes to the global network structure.

A popular model for visualizing the macroscopic connectivity structure of directed networks is the bow-tie model e.g., the bow-tie structure of the Web [40]. Figure 5 visualizes the bow-tie structure of the Legislation Network, while sizes of the various components are given in Table 4.

The bow-tie consists of the following components:

- SCC, The main (core) component is the connected component named SCC, containing all legal documents that can reach each other along directed edges,

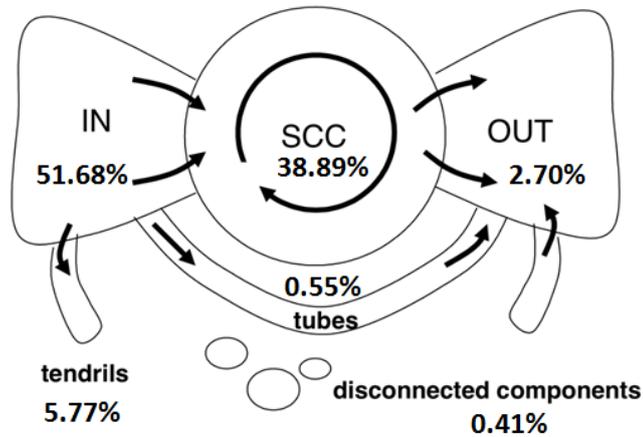

Fig. 4: Connectivity structure of the Legislation Network based on the Bow-tie model. One can pass from any node of IN through SCC to any node of OUT. Hanging off IN and OUT are TENDRILS containing nodes that are reachable from portions of IN, or that can reach portions of OUT, without passage through SCC. It is possible for a TENDRIL hanging off from IN to be hooked into a TENDRIL leading into OUT, forming a TUBE – a passage from a portion of IN to a portion of OUT without touching SCC.

Table 4: Sizes of bow-tie components for the Legislation Network

| Component | # of nodes | % of nodes |
|---|---|---|
| SCC | 91,107 | 38.89% |
| IN | 121,093 | 51.68% |
| OUT | 6,314 | 2.70% |
| TUBES | 1,300 | 0.55% |
| TENDRILS | 13,523 | 5.77% |
| OTHERS | 950 | 0.41% |

- IN, it contains non-core legal documents that can reach the core via a directed path,
- OUT, it consists of legal documents that can be reached from the core,
- TUBES, tubes are formed by non-core legal documents reachable from IN and that can reach OUT,
- TENDRILS, legal documents reachable from IN, or that can reach OUT, but not belonging to the above components,
- DISCONNECTED, the remaining documents are disconnected.

Observing the macro structure in the legislation network we note that the SCC (core) and IN components are larger and the OUT component significantly smaller, compared to other studies e.g, the web [40]. Currently, almost all nodes belong to the g.c. (0.41% disconnected component), about 40% of the nodes belong to the larger strongly connected component of the legislation network and 50% of nodes can reach the CORE with a direct path. The link structure of the legislation network is well interconnected. Most legal documents belong to the giant component, and from any document it is possible to reach almost any other. This is probably due to an implicit aim of the legislation system, that is driving to related/ connected documents. In this way the content of each document can be fully understood after visiting many different documents. Furthermore, observing the evolution of this macro structure in the Legislation Network, as can be seen in Figure 6, both the SCC (core) and G.C. components are getting larger over time.

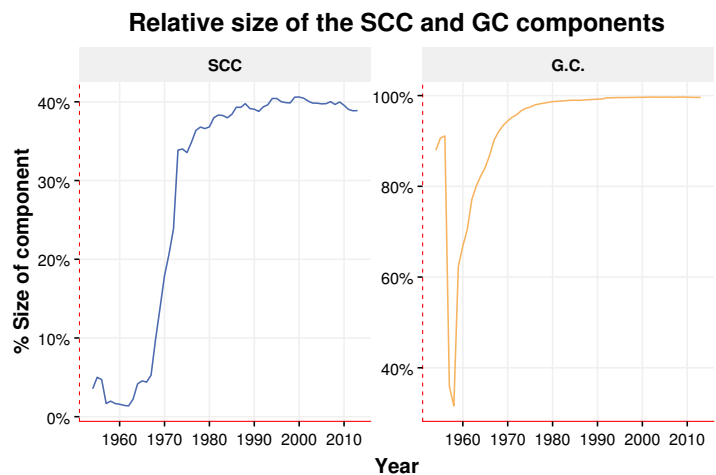

Fig. 5: Relative size of the SCC (core) and G.C. components with respect to the rest of the graph

A well established metric for networks is the *degree distribution*, $P(k)$, giving the probability that a randomly selected node has k links. A popular visualization of the degree distribution is the Lorenz curve, a type of plot to measure inequality originally used in economics [41]. In a network degree plot, the Lorenz curve is a straight diagonal line when all nodes have the same degree and curved otherwise. It visualizes statements of the form "X% of nodes with smallest degree account for Y% of edges". The Gini coefficient [42] is the ratio of the area between the line of perfect equality and the observed Lorenz curve to the area between the line of perfect equality and the line of perfect inequality. The Gini coefficient is mainly used in economics to characterize the inequality present in the distribution of wealth, but it can be used to measure the heterogeneity of any empirical distribution. The Gini coefficient takes values between zero and one, with zero denoting total equality between degrees, and one denoting the dominance of a single node. The higher the coefficient, the more unequal the distribution is.

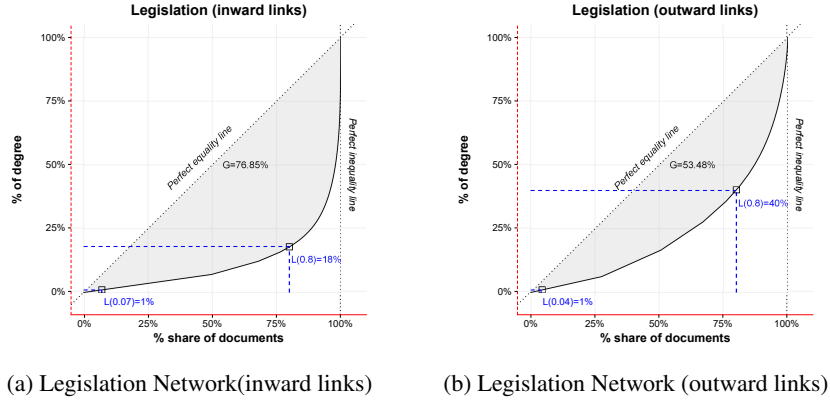

(a) Legislation Network(inward links)        (b) Legislation Network (outward links)

Fig. 6: Lorenz curve and Gini coefficient for the number of references per legal document distribution in the Legislation Network. We annotated values for the 1% effect and the Pareto 80/20 rule.(Best viewed in color.)

Figure 7 presents the Lorenz curve along with the Gini coefficient for the Legislation Network. If the Legislation Network were to be a random network, the Lorenz curve should be close to the straight diagonal line and its degrees should follow a Poisson distribution. Interestingly, in the contained plots, we can see that the Lorenz curve is not close to the straight diagonal line, but deviates towards inequality. The majority of documents are cross referenced by only a few times, while there are a few documents that are widely linked. We notice that numerous small-degree nodes coexist with a few hubs, nodes with an exceptionally large number of links. The top-1% of the highest degree nodes accounts for 7% of all inward and 4% of all outward links. We also annotated the Pareto 80/20 rule: percent of the highest degree nodes accounting for 80% of all links. For instance in Figure 7a we can see that 80% of all inward links is attributed to the 17% highest in-degree nodes in the Legislation Network, while 80% of all outward links is attributed to the 40% of the highest out-degree nodes , as showed in Figure 7b.

In many cases the *degree distribution*, $P(k)$, decays as a power-law, following

$$P(k) \infty K^{-\gamma} \tag{1}$$

where $\gamma$ is a constant parameter of the distribution known as the exponent or scaling parameter, that typically lies in the range $2 < \gamma < 3$. This feature is common to large scale communication, biological and social systems [5,22,43] and to the network of US Supreme Court decisions [1,3]. In practice, few empirical phenomena obey power laws for all values of $x$. More often, the power law applies only for values greater than some minimum $x_{min}$. In such cases, we say that the tail of the distribution follows a power law.

In order to classify the degree distribution in the Legislation Network, we fit a power law model to the degree distribution using the methodology outlined in [44]. Specifi-cally, each observed vertex degree, $x$, in the Legislation Network represents a candidate threshold value, $x_{min}$, above which the scaling behavior associated with the power law model may provide a plausible fit. We then find the associated parameter values which maximize the fit to the degree distribution above each candidate thresh-old value and select the best overall fit, given the corresponding threshold and parameter values. Testing whether the model is plausible given the network degree distribution, involves synthe-sizing $m = 2500$ sample degree distributions from a theoretical version of each model with the threshold and parameter values equal to those estimated for the observed de-gree distribution [10]. The power law model can be ruled out if 10%, $p < 0.10$, or fewer of the fits to the synthetic sets are poorer than the best fit to the actual data.

   In Table 5 we show results from fitting a power-law form to the Legislation Net-work, inward and outward links, alongside with a variety of generic statistics for the degree distributions such as mean, standard deviation, and maximum value. The last column of the table reports the $p-value$ for the power-law model, which gives a mea-sure of how plausible the power law is as a fit to the data [11].

Table 5: Basic statistics for the degree distribution such as mean, standard deviation, and maximum value of the Legislation Network, along with power-law fits. $n$ denotes the size of nodes, $n, tail$ is the size of the fitted power-law region, $\gamma$ is the scaling parameter and $x_{min}$ the restricted power-law fit. $p-value$ estimates were derived from a bootstrap using 2500 replications. (Statistically significant values are denoted in **bold**.)

| Network (Direction) | $n$ | $\langle x \rangle$ | $\sigma$ | $x_{max}$ | $x_{min}$ | $\gamma$ | $n_{tail}$ | $p$ |
|---|---|---|---|---|---|---|---|---|
| Legislation Network (In) | 141,798 | 7.04 | 54.91 | 6373 | $34 \pm 15$ | $2.24 \pm 0.15$ | $4351 \pm 29,700$ | **0.232** |
| Legislation Network (Out) | 224,856 | 4.44 | 8.2 | 1160 | $34 \pm 11$ | $3.55 \pm 0.42$ | $2210 \pm 36,000$ | **0.246** |

   Results of our heavy tailed analysis are plotted in Figure 8. The left column accounts for inward links while the right column represents outward links. For each sub-network, we provide the frequency distribution plot on log log scales (top row) and the cumu-lative degree distribution alongside with the fitted power-law and distributions (bottom row). We note that $\sigma$ values are significantly larger than $\langle x \rangle$, revealing large vari-ations in node degrees. The majority of documents are cross referenced by only a few times, while there are a few documents that are widely linked. Power law is a plau-sible fit for the distribution of inward links, $\gamma_{in} = 2.24$ and outward links, $\gamma_{out} = 3.55$ in the Legislation Network, with statistically significant $p-values$, 0.232 and 0.246 respectively.

   Useful insights to understand the origin of this heterogeneity, can be derived by the preferential attachment process[12], which was initially proposed in the context of wealth distributions [46] and afterwards utilized in complex networks [47]. The under-

---

[10] For the p-value to be accurate to about 2 decimal digits, we would choose to generate $m = 2500$ synthetic sets, as suggested in [44].

[11] We used the R poweRlaw package for heavy tailed distributions presented in [45].

[12] Also known as cumulative advantage or "the rich get richer".

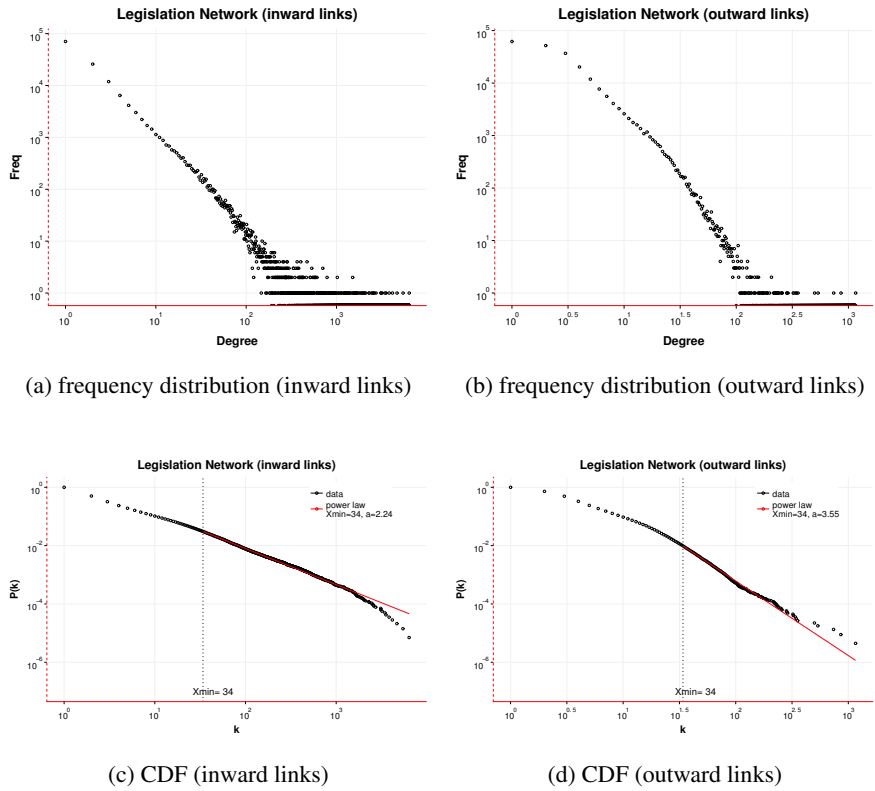

(a) frequency distribution (inward links)

(b) frequency distribution (outward links)

(c) CDF (inward links)

(d) CDF (outward links)

Fig. 7: The Legislation Network frequency distribution plot on log log scales for inward links 8a and outward links 8b. Fitted power-law and the cumulative degree distribution for inward links 8c and outward links 8d.

ling principle of the preferential attachment process is that new nodes attach preferentially to already well connected nodes, thus resulting in networks with skewed degree distributions. In such model, the distribution of connections is highly susceptible to its initial starting conditions. In the microscopic statistics of the legal domain when judges write opinions, they cite cases and other authorities that are the most relevant ones to the case they are deciding. With computer-assisted legal research tasks and widespread commercial legal search engines they are more likely to utilize higher-ranked legal document. Lawyers also rely on this cross citation as to form a well-grounded legal analysis. Therefore judges and lawyers are more likely to utilize or to link to, thus increasing the degree of, documents that already have a high degree (popular). Similarly, legal documents atop centrality-based rankings are there due to their high degree. If a legal document is already popular, it is more likely to receive another link. In other words, we notice a rich-get-richer phenomenon that amplifies the popularity of highly ranked documents.

Another important topological characteristic that many real graphs were found to exhibit is the so called *small-world* [29]. According to [32] small-world networks are defined as having a small diameter and high clustering. Many social, technological, biological and information networks have been studied and categorized as small-world networks [30]. Small-world networks can be seen as systems that are both globally and locally efficient, in terms of how efficiently information is exchanged over the network [48].

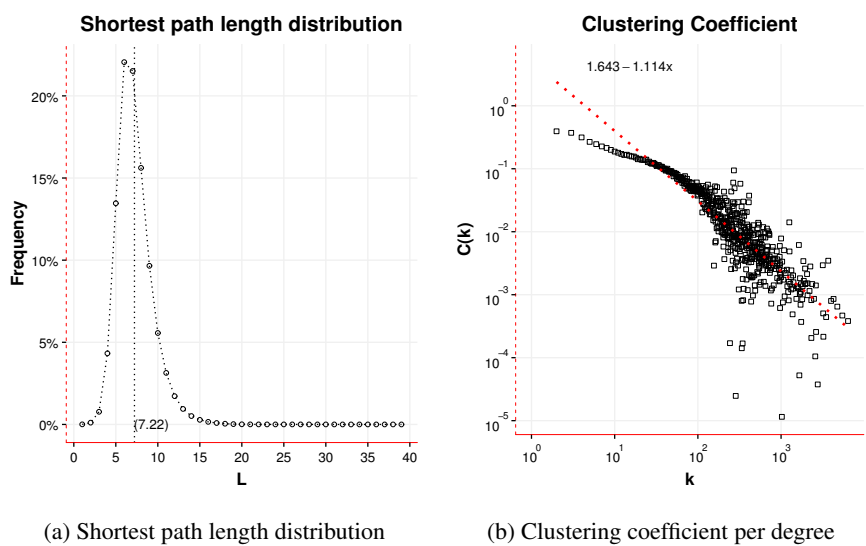

(a) Shortest path length distribution     (b) Clustering coefficient per degree

Fig. 8: a) The plot of shortest path length distribution for the Legislation Network. On an average each legal document is 7 hops away from any other in the network. b) Clustering coefficient per degree on log log scale. Slope of dotted line is $-1.1$ (Best viewed in color.)

Small world properties are measured by the average shortest path [13] and clustering coefficient [14] metrics. The distribution of shortest path lengths is plotted in Figure 9a. The diameter $D$ of the Legislation Network, defined as the maximum of the shortest path lengths, is 39 and the average shortest path, the mean of geodesic distance between any pairs that have at least a path connecting them, is 7.22. We do notice the presence of the "six degrees of separation" phenomenon in spite of huge size of the Legislation Network.

---

[13] Average number of steps along the shortest paths for all possible pairs of network nodes.

[14] Clustering coefficient is a measure of the degree to which nodes in a graph tend to cluster together.

The distribution of the degree-dependent clustering coefficient $C(k)$ is shown in 9b. For clarity, we added the line with slope $-1.1$ in the log-log scale. Although a clear power law may not be a plausible fit, the clustering coefficient is inversely proportional to degree $k$. High-degree nodes are linked to many nodes, probably belonging to different groups, thus, resulting in small clustering coefficient of the large-degree nodes. On the contrary, low-degree nodes generally belong to well-interconnected communities, corresponding to high clustering coefficient of the low-connectivity nodes. This pattern as studied in [49], indicates the existence of a hierarchical architecture in the network. A hierarchical architecture implies that sparsely connected nodes are part of highly clustered areas, with communication between the different highly clustered neighborhoods being maintained by a few hubs.

In order to classify a network as a small-world network, the candidate network metrics are compared with Erdös-Rényi random networks [31], with the same number of nodes and edges. If a network exposes the small world properties, then it is expected that average shortest path is slightly shorter than of a random network and the average clustering coefficient is of magnitude larger than that of a random network.

Similar to many studies on the small-world networks [30], our analysis is restricted to the giant components in the networks i.e., the maximal connected sub-graph of the network. Table 6 summarizes the results of our analysis on the Legislation Network and the current (active) versions of the legislation sub-networks we consider. Average shortest path and average clustering coefficient metrics are denoted by $L_{net}$ and $C_{net}$ and the corresponding random network ones are symbolized as $L_{rand}$ and $C_{rand}$ respectively.

Table 6: Small world metrics

|  | $L_{net}$ | $C_{net}$ | $L_{rand}$ | $C_{rand}$ |
|---|---|---|---|---|
| Legislation Network (LN) | 7,22 | 0,011 | 8,64 | 3,73e-05 |
| (current) Legislation (LN) | 7,58 | 0,0215 | 7,97 | 7,86e-05 |
| (current) Regulations (RN) | 7,2 | 0,00707 | 12,4 | 0,000156 |
| (current) Inst. cited (ICN) | 6,9 | 0,0343 | 7,26 | 0,000126 |
| (current) Legal basis (LBN) | 1,48 | 0,000278 | 24,1 | 5,68e-05 |

We notice that, despite the variations in the metrics, all of the networks satisfy the small-world conditions. Interestingly not only the Legislation Network presents small world characteristics, but also the current (active) versions of the Legislation sub-networks also. Comparing our results with other studies, as presented in [30], we see that the average shortest path lengths in the legislation sub graphs are distinctively smaller than the values of networks reported and of magnitude smaller than the theoretical average degree of the corresponding random model. We attribute this finding to the nature of law and its hierarchical form. Legal documents are made by the authority given by other legal documents, which reduces their total number of references well below the expected number from the random model.

Finally, the tendency for nodes in networks to be connected to other nodes that are like (or unlike) them in some way, assortativity, has been studied in [50]. While assortativity is often examined in terms of a node's degree, other discrete criteria/ characteristics of the examined network models can be applied. In terms of degree values, social networks, exhibit assortative mixing since nodes tend to be connected with other nodes with similar degree values, while on the contrary technological and biological networks exhibit dissortativity, as high degree nodes tend to attach to low degree ones. The level of assortative mixing in a network is measured by the assortativity coefficient, with positive values implying that the network is assortative, negative values that it is dis-assortative and zero assortativity shows no correlation.

The Legislation Network exhibits a small degree of dis-assortativity (-0.0904) in terms of nodes degree, as high degree nodes (hubs) are more likely to connect to nodes of lower degree. On the contrary, the assortativity coefficient in terms of document type connectivity, sector classification Table 1, is 0.443, revealing that there is a strong tendency of legal documents to connect with legal documents belonging to the same type and thus form clusters of the same sector.

### 4.2 Temporal Evolution of Legislation

Real-world networks evolve over time by the addition and deletion of nodes and edges. The Legislation Network also evolves over time, with nodes and edges appearing or disappearing, as new laws are being continuously created and other laws are amended, invalidated or cease to exist. A complementary issue, over-looked in the legal citation network literature, are the temporal aspects of those networks. As with network topology, the temporal structure of node/ edge activation's can affect dynamics of systems interacting through the network [51].

In this section, we present our main findings on studying the evolution of legislation over time. We analyze the temporal evolution of the Legislation Network by deriving static graphs that capture both temporal and topological properties of the system. We segment the legislation data into adjacent time windows, annually divided time intervals, considering only active nodes and edges, and then study the time evolution of the network structure in these windows.

In order to evaluate the temporal evolution of the Legislation Network, we considered sub-graphs at annually divided time intervals (*time step frames* $- t = 1 year$). For each time frame, from 1951 up to 2013, we create a sub-network using all legislation that it was in effect on year $Y$, by incrementally removing nodes/ edges, legal documents/ relations, from the legislation graph, according to the dates of affect and cancellation of affect within the current time frame, utilizing Algorithm 3.

Figure 10a illustrates the respective evolution on a sector basis. The x axis represents time while the y axis represents the growth of legislation corpus per sector. The active legislation corpus grows over the years in respect to all types of sectors. Similarly, in Figure 10b , the y axis represents the growth of legislation corpus along the various edges of the legislation graph. The number of active edges, connection between active legal documents, grows over the years in respect to all types of reference categories.

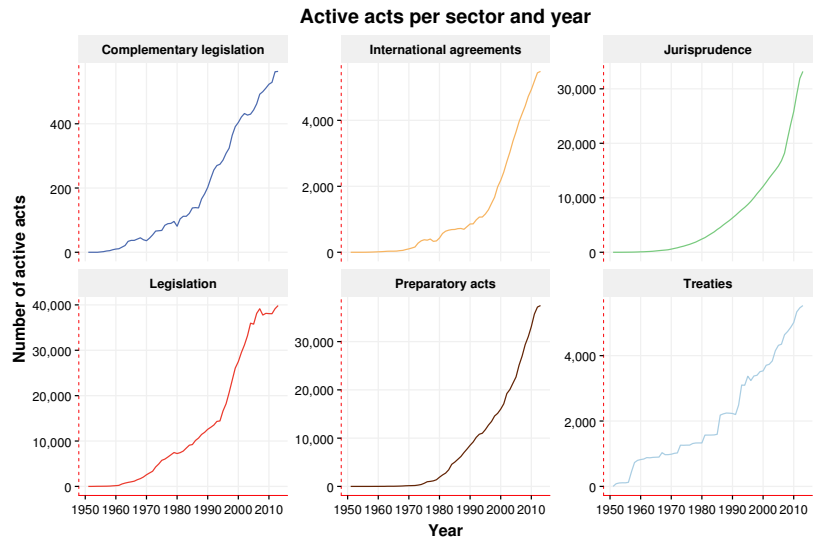

(a) Active Legislation per sector and year in the EU

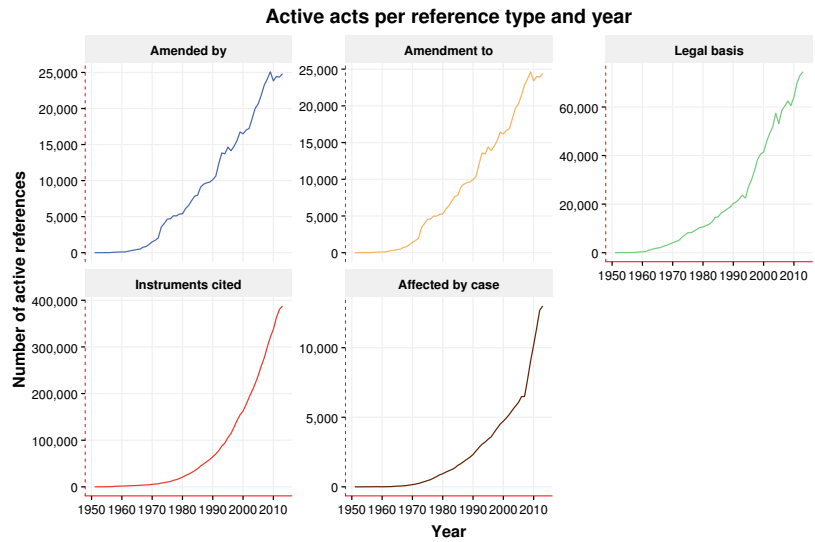

(b) Active Legislation per reference type and year in the EU

Fig. 9: a) Active Legislation per sector and year in the EU, (b) Active Legislation per reference type and year in the EU. The active legislation corpus is steadily growing over all types of sectors and reference categories. (Best viewed in color.)

Additionally, we also examined the evolution of these graphs over time. Leskovek et al. [6] studied a range of different networks, from several domains, focusing on the way in which fundamental network properties vary with time. They concluded that the densification power law is a property that holds across a range of diverse networks. According to the densification power law the number of edges is growing faster than the number of nodes.

In general, the densification power law is defined from the following form:

$$E(t) \infty N(t)^{\alpha} \tag{2}$$

where $E(t)$ and $N(t)$ denote the number of edges and nodes of the graph at time $t$, while $a$ ranges between 1 and 2.

In detail, our analysis was conducted on the four representative sub-networks presented in the previous section. Furthermore, the current version of each sub-network was formed and analysed for each year over our Legislation Network. In accordance with the findings of [6], the Legislation Network also follows a densification power law; the number of edges, connection between legal documents, grow faster than the number of nodes, legal documents[15]. Results of our temporal analysis are presented in Figure 11. For each sub-network, we illustrate on log log scales the number of active edges (relations between legal documents) and nodes (legal documents). The number of active relations between legal documents is growing faster than the number of active legal documents for all the sub-networks examined. We plan on a future work to utilize this property as to provide an evolutionary model, in order to describe the legislation process.

### 4.3   Resilience of the Legislation Network

While myriads of regulations try to formalize and regulate systemic risk in various domains e.g., financial systemic risk, the concept of systemic risk applies to every complex system. Cascading failures in a network of interconnected system components has been studied in the literature [52], but overlooked in legal domain. As stated in [53] the legal system must not only anticipate systemic failures in the systems it is designed to regulate, but also anticipate systemic risk in the legal system as well. Motivated by this idea, within this section, we describe an experiment for further studying the resilience of the Legislation Network. We do acknowledge the underling complexity of such evaluation that depends on legal experts explaining the legal consequences of the raw outcomes.

A fundamental issue in the analysis of complex networks is the assessment of their stability, aiming to understand and predict the behavior of a system under any type of malfunctions. Resilience refers to the ability of a network to avoid breakdowns when a fraction of its components is removed. Over the past few years a large number of networks have been evaluated for tolerance to errors and attacks, while several approaches have been proposed [23,54,30]. Usually tolerance to errors is measured in terms of changes to the diameter or the size of the giant component of the networks under evaluation when a fraction of nodes are removed in a random manner. On the contrary, under the assumption that an malicious agent will deliberately target the most connected nodes, tolerance to attacks is measured when a fraction of the most connected nodes, sorted in decreasing order, are removed from the network.

---

[15] We note that our findings, overall expansion (growth) of the network, are based on examining the active legislation network in each adjacent time window and not on the whole "static" version of the Legislation network. In the latter case this is an obvious finding since legal documents cite previous legal documents.

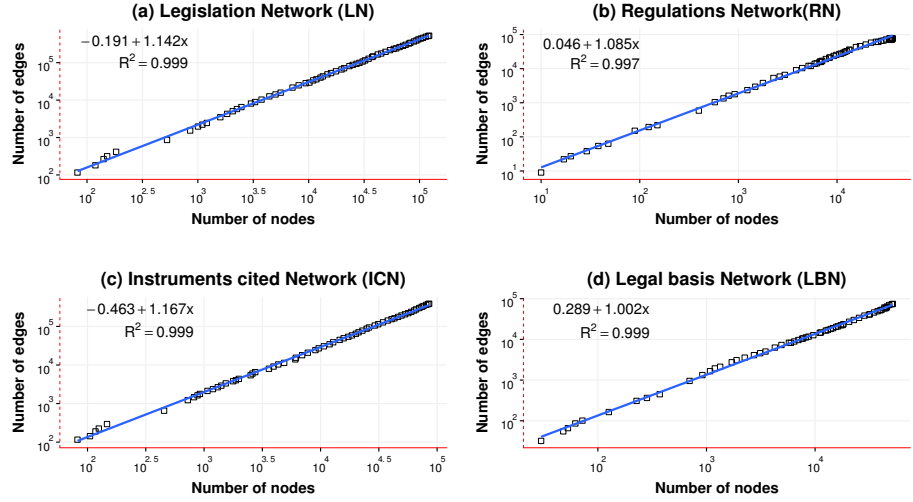

Fig. 10: Number of edges $e(t)$ versus number of nodes $n(t)$, in log-log scales, for the active version of legislation sub-graphs. All graphs obey the Densification Power Law, with a consistently good fit. Slopes: a = 1.142, 1.085, 1.167 and 1.002, respectively. Values for the coefficient of determination $R^2$ are also reported.

In our experimentation, we analysed and quantitatively measured the behavior of the Legislation Network, in case where some of its nodes (legal documents) or edges (connections) between nodes are removed. In real-life cases, this may be reflected when laws are amended, invalidated or cease to exist, especially during major deregulation reforms into various industry segments. Furthermore complying with one rule could require actions that make complying with another rule more difficult. Similarly, because legal rules often are interrelated through techniques, such as cross-referencing and stare decisis, the way one rule is interpreted and applied could affect the meaning or operation of other rules [53].

We evaluated the changes in giant component of the graph, which is the largest connected sub-graph, when a small fraction of the nodes is removed [16]. Since we cannot think of a real life scenario, where inactive legislation get's invalidated, we use in our analysis the current (active) version of each sub-network as presented in Section 3.3, utilizing Algorithm 3. In order to simulate errors we randomly removed nodes, while for simulating attacks we removed nodes according to their degree in decreasing order.

---

[16] The deletion of a node causes also the deletion of all of its edges.

For each of the four legislation sub-graphs, we have created an Erdös-Rényi random network with same number of nodes and edges. Those random networks helped us to visualize the effects of power law distribution and small-world properties that were previously described. All the eight networks were tested under our error/ attack assumptions with a removal rate of 5% of remaining nodes at each step. Then, on each step, we calculated the giant component of the network according to the amount of remaining nodes. The whole procedure was repeated $1,000$ times and averaged values of the fraction of nodes in the giant component were calculated.

Results of our resilience evaluation are presented in Figure 12. For each sub-network, we illustrate the percentage of the giant component according to the fraction of removed nodes. According with simillar studies in the literature [5], the Legislation Network, a scale-free network, presents an exceptional robustness against random node failures. However, as a result of this resilience, in cases where the highest number of edges are attacked, the Legislation Network breaks down earlier than random networks. The Legislation Network (LN) is a scale-free network with many low degree nodes and a few highly connected ones. Random node removal affects mostly low degree nodes, thus marginally altering the network topology. In such cases, the network behaves like a random network. At the same time, the removal of the highly connected nodes has a catastrophic effect leaving the network highly divided.

As far as the Instruments Cited (ICN) sub-network, which resembles a citation network, appears to be the most resilient among the others. However, using it in such scenarios would provide us with inaccurate results, since simple citations do not often convey any special meaning in the legislation process.

On the other hand, the Legal Basis(LBN) legislation sub-graph appears to be the least resilient of the four. This sub-network consists only of edges of type Legal basis and it represents the internal hierarchy of the legislation corpus. Highly influential laws, which serve as legal basis for several others, play an important role in the Legislation Network. These laws keep the network connected and modifications on them can induce serious consequences on the Legislation Network.

For a real life example, of the consequences that might occur after a single court decision we consider Ireland's drug loophole case[17]. Ireland's Court of Appeal found parts of the 1977 Misuse of Drugs Act to be unconstitutional [18], since the act was added via ministerial order and without consulting the Oireachtas (both houses of the Irish parliament). After the Act was first passed, almost 40 years ago, successive amendments have added drugs to the original "banned" list. With the court's decision all off the amendments have been declared invalid, thus, opening a loophole in the legal system. In accordance with our previous findings on the structure and the topology of the Legislation Network, e.g., scale free network, small world, growing diameter and densification power law, we expect that this type of legal "accidents" will become more frequently and with exponential cascading failures.

Authorities at various levels within the E.U. e.g., European, national, local level would benefit from an impact analysis of changes to the Legislation Network. Our pro-posed model should provide assistance to legal experts, as to properly access the legal consequences of proposed legislation changes.

---

[17] edition.cnn.com/2015/03/11/europe/ireland-legal-drugs/

[18] Stanislav Bederev v Ireland, The Attorney General and the Director of Public Prosecutions (Irish Court of Appeal), 1409 (2014).

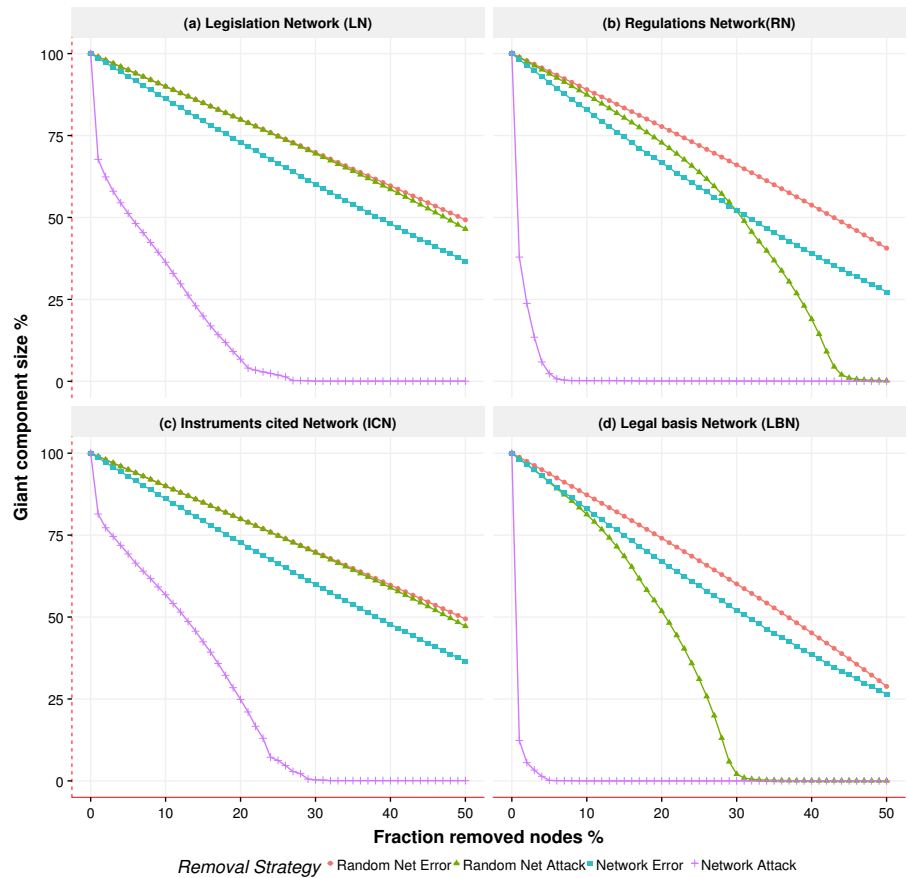

Fig. 11: Resilience of the Legislation Network. Fraction of nodes in the giant component as a function of the fraction of removed nodes in the 4 legislation sub-networks and random networks with the same dimensions. The Instruments Cited sub-network appears to be the most resilient of the four under targeted attack and the Legal Basis as the least resilient. (Best viewed in color.)

## 5   Conclusions

In this paper, we introduce a network-based approach to model the law: the Legislation Network. Our approach offers a model to create a systematic alternative structure to a naturally evolved normative system. The Legislation Network is a multi-relational network that accommodates the hierarchy between the sources of law and can represent relationships of various categories between legal documents, along with their temporal evolution.

To the best of our knowledge, this work significantly differs from most previous legal citation analysis studies, and the monolithic view to the legislation corpus they share. We assume that there exist multiple, heterogeneous legislation sub-networks and a sophisticated examination of their properties would generate important new relationships in the real-life paradigm.

Characterizing the structural properties of a network is of fundamental importance to understand the complex dynamics of the modelled system. The Legislation Network is highly heterogeneous with respect to the number of edges incident on a node. The degree distribution of legal documents follows a power law and, even it is resilient to the random loss of nodes, it is very vulnerable to attacks targeting the high-degree ones. The connectivity of the Legislation network relies on a small set of very important legal documents. Modifying such legal documents, like actions of amending or cancellation, can cause an avalanche of unintended consequences to the legislation corpus. We plan to further evaluate the resilience of the sub-networks by employing a wider range of criteria to determine the importance of the removed under attack nodes, such as betweenness, Hits and PageRank.

We also studied the temporal evolution of the Legislation Network. Results showed that the Legislation Network becomes denser over time, with the number of edges growing faster than the number of nodes. Further studies based upon the discovered characteristics of Legislation Network may provide us with a richer model to better explain the structure and evolution of legislation. Towards this, we plan to further evaluate whether the graph patterns observed in the current study can be fitted into other well established graph generators, like the Preferential attachment [5] and Forest Fire [6] models.

In parallel to all the above, we intend to extend our model and use it for link prediction, trying to predict whether and how many times a legal document will be cited in the future, given its position in the evolved Legislation Network. A more sophisticated approach will be to predict which legal documents (and/or when) will become amended or even invalidated.

In addition, our model can be exploited for visualizing the legal corpus. Graph visualizations are used to convey the content of a graph as they can highlight patterns, reveal clusters and related connections. We believe that a visualization system for the Legislation Network can be of great assistance to both citizens and legal experts, helping them to easily navigate the legislation corpus. Another great benefit of such an approach, lies in the fact that legislation can be exploited not only from the traditional point-of-view, but as a graph of hyper-textual information with temporal properties. As an example, it will be easier for lawmakers to monitor the effect of a possible change in the whole normative system, thus taking appropriate actions. In such a system, the use of domain-specific ontologies [55] and linked data techniques [56] would further enrich the added value of Legislation Network.

Finally, our modelling approach can be used to improve the effectiveness of legal information retrieval systems [57]. Our hypothesis is that the Legislation Network can be exploited for text retrieval, in the same manner as hyperlink graphs on the Web.